\documentclass[prd,nofootinbib,twocolumn,showpacs]{revtex4}
\usepackage{graphics}
\usepackage{bm}

\begin{document}

\title{Signatures of kinetic and magnetic helicity in the CMBR}

\author{Levon Pogosian}
\affiliation{Theoretical Physics, The Blackett Laboratory,
Imperial College, Prince Consort Road, London SW7 2BZ, United Kingdom.}

\author{Tanmay Vachaspati}
\affiliation{Department of Physics, Case Western Reserve University,
10900 Euclid Avenue, Cleveland, OH 44106-7079, USA.}

\author{Serge Winitzki}
\affiliation{Department of Physics and Astronomy,
Tufts University, Medford, MA 02155, USA.}

\begin{abstract}
P and CP violation in cosmology can be manifested as large-scale helical
velocity flows in the ambient plasma and as primordial helical magnetic
fields. We show that kinetic helicity at last scattering leads to
temperature-polarization correlations ($C_l^{TB}$ and $C_l^{EB}$) in
the cosmic microwave background radiation (CMBR) and calculate the
magnitude of the effect. Helical primordial magnetic fields, expected
from cosmic events such as electroweak baryogenesis, can lead
to helical velocity
flows and hence to non-vanishing correlations of the temperature and
B-type polarization. However we show that the magnitude of the induced
helical flow is unobservably small because the helical component of a
magnetic field is almost force-free. We discuss an alternate scheme for
extracting the helicity of a stochastically homogeneous and isotropic
primordial magnetic field using observations of the CMBR.
The scheme involves constructing Faraday rotation
measure maps of the CMBR and thus determining the sum of the helical and
non-helical components of the primordial magnetic field. The
power spectrum of B-type
polarization fluctuations, on the other hand, are sensitive only to the
non-helical component of the primordial magnetic field. The primordial
magnetic helicity can then be derived by combining these two sets of
observations.
\end{abstract}


\

\maketitle

The cosmic microwave background radiation (CMBR) has become
a remarkable tool for mapping the past and present states of the
universe. As more refined observations are made, a larger array
of theoretical ideas will be put to the test and more details of
the history of the universe will emerge.

Until now, only the temperature anisotropy of the CMBR has been
measured. It is widely recognized that even the most precise
measurement of the temperature spectrum would still be compatible
with a large class of initial conditions for the fluctuations in
the primordial universe. Measurements of the CMBR polarization
would significantly reduce this degeneracy. The much anticipated
first detection of the CMBR polarization could be provided by the
MAP satellite, launched by NASA in June of 2001, or balloon-borne B2K and
MAXIPOL experiments. There are several other ongoing or
planned ground-based experiments that have the detection of CMB polarization
as their primary goal. However, most expectations of obtaining polarization
spectra are connected with the Planck satellite mission,
currently scheduled to be launched in 2007.

A possibility that has already received some attention is that
large-scale Parity (P) violation may be observed via the CMBR
\cite{ScaFer97,LueWanKam99}. In Ref.~\cite{ScaFer97} it was shown
that a coherent magnetic field would induce non-zero P-violating
correlations in the CMBR through Faraday rotation. In
Ref.~\cite{LueWanKam99} the P violation was due to the dynamics of a
pseudoscalar field which could induce a fixed-handed rotation of
the polarization plane of the CMBR. In another case studied
in Ref.~\cite{LueWanKam99}, the pseudoscalar field could
produce an excess of left- over right-handed polarization states of
gravitational waves, which would be manifested in the CMBR
as a P-violating signature. In Sec.~\ref{hvel} we examine the
consequence of yet another possible source of P-violation
(and also CP-violation), namely large-scale helical velocity flows
({\it i.e.}~kinetic helicity) at recombination.
We find that the CMBR temperature and polarization cross-correlators
can provide a measurement of kinetic helicity (Sec.~\ref{hvelcmbr}).
A motivation for considering helical flows is that helical
magnetic fields may have been generated in the early universe
\cite{FieCar00,Cor97,Vac01a,Vac01b} and our initial guess was
that such fields could induce helical velocities. However,
in Sec.~\ref{hvelbhel} we show that the magnitude of the
kinetic helicity induced by helical magnetic fields is
insignificant and does not lead to any observable signature in
the CMBR.

This leaves us with the challenge of devising a methodology
to observe the helicity of primordial magnetic fields.
In Sec.~\ref{strategy} we describe a method
which involves observing the temperature and polarization
fluctuations of the CMBR and also the Faraday
rotation measure of the CMBR polarization along different
directions. From a suitable combination of
these observations, it is possible to extract the
primordial magnetic field helicity. Hence, at least
in principle, the helicity of cosmic magnetic fields may be
observable.

\section{Kinetic helicity}
\label{hvel}

Kinetic helicity of a velocity field ${\bf v}$
is characterized by a non-vanishing
value of $\langle {\bf v}\cdot {\bm \omega}\rangle$, where
${\bm \omega}\equiv \nabla\times {\bf v}$. For the purpose
of calculating the effect of helical flows on the cosmic
microwave background radiation, it is useful to decompose
the velocity field at last scattering into gradient
components $v^{(0)}$ and
rotational components $v^{(\pm 1)}$ as follows:
\begin{eqnarray}
\label{eq:vdecomp} \nonumber
v_{j}\left( {\mathbf x} \right) &=& \int {d^{3}k \over (2\pi)^3}\,
e^{-i{\mathbf k}\cdot {\mathbf x}} \\
& \times & \left( i\hat{k}_{j} v^{(0)}+Q_{j}^{(1)}v^{(1)}+
Q_{j}^{(-1)}v^{(-1)}\right) ,
\end{eqnarray}
where
\begin{equation}
{\mathbf Q}^{(\pm 1)}\left( {\mathbf k}\right) \equiv -i\frac{\hat{{\mathbf
e}}_{1}\pm i\hat{{\mathbf e}}_{2}}{\sqrt{2}}
\end{equation}
and the vectors $\hat{{\mathbf e}}_{1,2}$ are such functions
of ${\mathbf k}$ that they form an orthonormal basis together
with $\hat{{\mathbf e}}_{3}\equiv \hat{{\mathbf k}}$. The functions
${\mathbf Q}^{(\pm 1)}$ can be thought of as eigenvectors of helicity because
$i\hat{{\mathbf k}}\times {\mathbf Q}^{(s)}=s{\mathbf Q}^{(s)}$, where $s=\pm 1$.
Also, ${\mathbf k}\cdot {\mathbf Q}^{(s)}=0$ and ${\mathbf Q}^{(s)}\left( {\mathbf k}\right)
={\mathbf Q}^{(-s)}\left( -{\mathbf k}\right) $. Under a parity transformation,
${\mathbf v}({\mathbf x})\rightarrow -{\mathbf v}({\mathbf {-x}})$,
${\mathbf k}\rightarrow -{\mathbf k}$ while ${\mathbf Q}^{(\pm 1)}\rightarrow {\mathbf Q}^{(\mp
1)}$ and therefore $v^{(0)}\left( {\mathbf k}\right) \rightarrow
-v^{(0)}\left( -{\mathbf k}\right) $, $v^{(\pm 1)}\left( {\mathbf k}\right)
\rightarrow -v^{(\mp 1)}\left( -{\mathbf k}\right) $.

In Appendix~\ref{appendixA} we will show that the parity-odd CMB correlators
$C_{l}^{TB}$ and $C_{l}^{EB}$ are linearly dependent on the expectation
value of the parity-odd combination of $v^{(\pm 1)}$:
\begin{equation}
\langle v^{(1)}\left( {\mathbf k}\right) v^{(1)}\left( -{\mathbf k}\right)
-v^{(-1)}\left( {\mathbf k}\right) v^{(-1)}
\left( -{\mathbf k}\right)\rangle \, .
\label{eq:vv}
\end{equation}
The average helicity of the velocity field is proportional to the same
quadratic combination of $v^{(\pm 1)}$. Namely,
\begin{eqnarray}
&&\int d^{3}{\bf x} ~ {\mathbf v}( {\mathbf x}+{\mathbf y} )
\cdot [\nabla \times {\mathbf v} ( {\mathbf x})] =
\int {d^{3}{\mathbf k} \over (2\pi)^3}
k e^{i{\mathbf k}\cdot {\mathbf y}}
\nonumber \\
&&\times \left( v^{(1)}( {\mathbf k}) v^{(1)}( -{\mathbf k}) -
v^{(-1)}( {\mathbf k}) v^{(-1)}( -{\mathbf k}) \right),
\end{eqnarray}
where $k\equiv |{\bf k}|$.

Let us assume that the initial velocity field is random with a
given power spectrum, such as
\begin{equation}
\left\langle v_{i}( {\mathbf k}) v_{j}( {\mathbf k}') \right\rangle =
v_{0}^{2} \frac{k^n}{k_{*}^{n+3}} (2\pi)^3
\delta \left( {\mathbf k}+{\mathbf k}'\right) P_{ij}( \hat{{\mathbf k}}) ,
\label{vikvjk}
\end{equation}
where $v_0$ is the characteristic velocity of the flow and $k_*$ is the wave
vector corresponding to a cutoff scale. In general, $P_{ij}$ does not have to
be a symmetric tensor. The restrictions on it are: the reality condition
$P^{*}_{ij}( \hat{{\mathbf k}}) = P_{ij}(-\hat{{\mathbf k}}) = P_{ji}(
\hat{{\mathbf k}})$, the divergence-less condition (incompressibility)
$k^{i}P_{ij}=0$, and the
requirement of correct transformation under rotations, which forces $P_{ij}(
\hat{{\mathbf k}})$ to be a tensorial function of $\hat{k}^i$. The latter
condition is needed to ensure that the correlator $P_{ij}(\hat{{\mathbf k}})$
describes a homogeneous and isotropic random vector field. It follows from
these restrictions that
\begin{equation}
\label{eq:Pij}
P_{ij}( \hat{{\mathbf k}}) =
f(k) \left[ \delta _{ij}-\hat{k}_{i}\hat{k}_{j}\right] +
ig(k) \varepsilon_{ijl}\hat{k}^{l},
\label{Pijfg}
\end{equation}
where $f(k)$ and $g(k) $ are real functions of $k=|{\bf k}|$.
By using
$$
\left \langle
|{\bm \omega}({\bf k}) \pm k {\bf v}({\bf k})|^2
\right \rangle
\ge 0,
$$
where ${\bm \omega}({\bf k})$ is the Fourier transform of
${\bm \omega}({\bf x})$,
we can deduce that the functions $f$ and $g$ are not completely
independent but must satisfy the inequality
\begin{equation}
f({\bf k}) \ge |g({\bf k})| \, .
\label{maxkinhel}
\end{equation}
[$f({\bf k}) \ge 0$ follows by taking the trace of
Eq.~(\ref{vikvjk}).] Eqs.~(\ref{eq:vdecomp}) and (\ref{vikvjk})
now lead to
\begin{eqnarray}
\left\langle v^{(1)}\left( {\mathbf k}\right) v^{(1)}\left( {\mathbf
k}'\right) -v^{(-1)}\left( {\mathbf k}\right) v^{(-1)}\left( {\mathbf
k}'\right) \right\rangle
\nonumber \\
= v_{0}^{2} \frac{k^n}{k_{*}^{n+3}} (2\pi)^3
\delta \left( {\mathbf k}+{\mathbf k}'\right)
P^{ij}( \hat{{\mathbf k}})
Q_{[i}^{(1)}( \hat{{\mathbf k}})
Q_{j]}^{(-1)}( \hat{{\mathbf k}}) .
\end{eqnarray}
The $f$-term in Eq.~(\ref{Pijfg}) will not produce helicity
because it is symmetric in $i$, $j$. Since
\begin{equation}
i\varepsilon _{ijl}\hat{k}^{l}Q_{[i}^{(1)}( \hat{{\mathbf k}})
Q_{j]}^{(-1)}( \hat{{\mathbf k}}) =1,
\end{equation}
we get
\begin{eqnarray}
\label{odd_{s}pectrum}
\left\langle v^{(1)}( {\mathbf k})v^{(1)}( {\mathbf k}') -
v^{(-1)}( {\mathbf k}) v^{(-1)}( {\mathbf k}') \right\rangle
\nonumber \\
= v_{0}^{2} \frac{k^n}{k_{*}^{n+3}}
(2\pi)^3 \delta \left( {\mathbf k}+{\mathbf k}'\right) g(k) .
\end{eqnarray}

\section{CMBR signature of kinetic helicity}
\label{hvelcmbr}

The CMBR temperature and polarization at a given direction on the
sky are described by time-averaged components of the electric
field intensity tensor $I_{ij} \equiv E_i E_j$. Equivalently, the
CMBR is represented by the Stokes parameters $I$, $Q$, $U$ and $V$
\cite{Cha60,Kos96}, where $I$ is the total intensity,
$Q$ and $U$ are two components of the linear
polarization and $V$ quantifies the circular polarization.
Assuming that the CMB photons prior to the last scattering were
unpolarized, we can drop $V$ from consideration, since
Thomson scattering can only generate linear polarization.

\begin{figure}
\scalebox{0.420}{\includegraphics{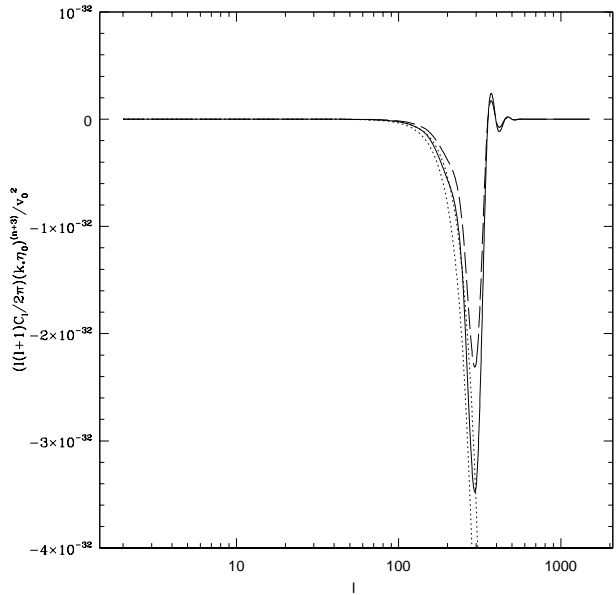}}
\caption{\label{fig:n2}
$(1/2\pi) l(l+1)C_l^{TB,EB}
(k_* \eta_0)^{n+3}/v_0^2
$  vs $ l $ for $ n=2 $ and
$2\pi / k_* = 288 $ Mpc, where $n$, $k_*$ and $v_0$ are defined in
Eq.~(\protect \ref{vikvjk}) and $\eta_0$ is the conformal time today.
We have also taken $g(k)=1$ [defined in Eq.~(\protect \ref{Pijfg})].
The solid line shows the $TB$ correlation and the dashed
line is $EB$. Dotted lines show the same spectra in the absence of a cutoff.}
\end{figure}

Decomposing the temperature anisotropy into spherical harmonics gives
\begin{equation}
\Delta\left( \hat{{\mathbf n}}\right) \equiv
{ T\left( \hat{{\mathbf n}}\right) - \bar{T} \over \bar{T}}
=\sum _{l,m}Y_{lm}\left( \hat{{\mathbf n}}\right) a^{T}_{lm}  \, .
\end{equation}
Analogously,
$Q\left( \hat{{\mathbf n}}\right) $ and $U\left(\hat{{\mathbf n}}\right)$
can be decomposed using spin-$2$ spherical harmonics (see
Refs.~\cite{ScaFer97,HuWhi97}):
\begin{equation}
\left( Q\pm iU\right) \left( \hat{{\mathbf n}}\right) =
\sum _{l,m}\, _{\pm 2}Y_{lm}a_{lm}^{\left( \pm 2\right) }.
\end{equation}
The ``electric'' and ``magnetic'' components of polarization
are eigenstates of parity and may be defined by
\begin{equation}
a_{lm}^{E}=-\frac{1}{2}\left( a_{lm}^{\left( 2\right) }+a_{lm}^{\left(
-2\right) }\right) ,\ a_{lm}^{B}=-\frac{1}{2i}\left( a_{lm}^{\left(
2\right) }-a_{lm}^{\left( -2\right) }\right) .
\end{equation}
Under parity inversion the components transform as
$a_{lm}^{T,E}\rightarrow \left( -1\right) ^{l}a^{T,E}_{lm}$, while
$a_{lm}^{B}\rightarrow -\left( -1\right) ^{l}a_{lm}^{B}$.

Observations of CMBR are usually presented in the form of
spectral functions $C_l^{X,Y}$ defined by
\begin{equation}
\label{clxy_definition}
C_{l}^{XY}=\frac{1}{2l+1}\sum _{m=-l}^{l}\left\langle \left(
a_{lm}^{X}\right) ^{*}a_{lm}^{Y}\right\rangle  \, ,
\end{equation}
where $X$ and $Y$ stand for $T$, $E$ or $B$.
Correlators in Eq.~(\ref{clxy_definition}) are real because
they involve summation over both positive and negative values of $m$.
This can be seen by noting that the
tensor spherical harmonics $_{s}Y_{lm}$ satisfy
\begin{equation}
\left[ _{s}Y_{lm}\right] ^{*}=\left( -1\right) ^{s+m}\, _{-s}Y_{l,-m} \, ,
\end{equation}
and therefore the components $a_{lm}^{\pm 2}$, $a_{lm}^{T,E,B}$ transform
under complex conjugation as
\begin{equation}
\left[ a_{lm}^{\pm 2}\right] ^{*}=\left( -1\right) ^{m}a_{l,-m}^{\mp
2}\, ,\ \left[ a_{lm}^{T,E,B}\right] ^{*}=\left( -1\right)
^{m}a_{l,-m}^{T,E,B}.
\end{equation}

The correlators $C_{l}^{TB}$ and $C_{l}^{EB}$ are parity-odd, while
all other correlators $C_{l}^{XY}$ are parity-even.
If no parity-odd sources are present, the ensemble
of fluctuations is statistically parity-symmetric and therefore the
correlators $C_{l}^{TB}$ and $C_{l}^{EB}$ must vanish. This is the
case in most of the literature on CMBR polarization. In our case,
however, the presence of parity-violating sources allows nonzero values
of the parity-odd correlators $C_{l}^{TB}$ and $C_{l}^{EB}$ and
hence these correlators are of special interest to us.

\begin{figure}
\scalebox{0.420}{\includegraphics{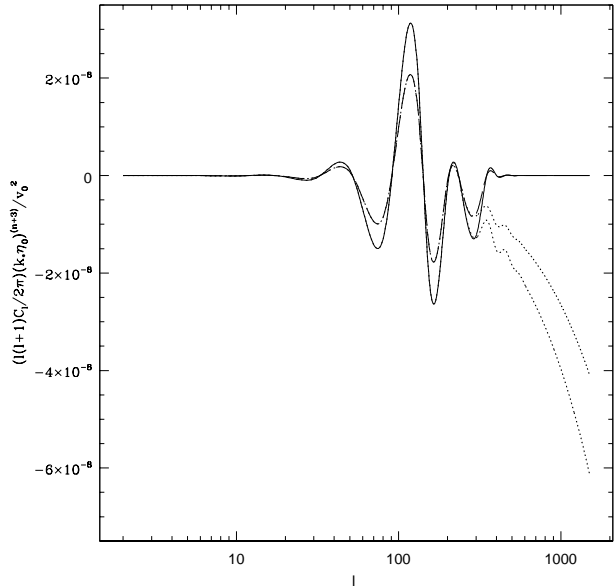}}
\caption{\label{fig:nm2} Same spectra as in Fig.~\protect \ref{fig:n2}
for $n=-2$ with (solid and dashed lines) and without (dotted lines)
presence of a cutoff at $2\pi/k_* = 288$ Mpc.}
\end{figure}

\begin{figure}
\scalebox{0.420}{\includegraphics{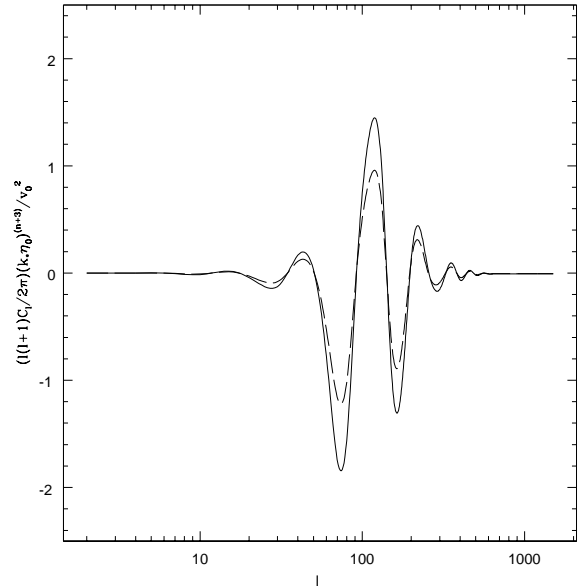}}
\caption{\label{fig:nm3} Same spectra as in Fig.~\protect \ref{fig:n2}
for $n=-3$. Spectra no longer depend on the cutoff scale.}
\end{figure}

The details of the calculation of $C_l^{TB}$ and $C_l^{EB}$
are given in Appendix \ref{appendixA}, where we find
some limiting forms of these functions. We have performed
numerical evaluations for several different values of $n$
and $k_*$ [defined in Eq.~(\ref{vikvjk})] and assuming $g(k)=1$
[defined in Eq.~(\ref{Pijfg})].
The results for $n=2$, $n=-2$ and $n=-3$ are shown in
Figs.~\ref{fig:n2}-\ref{fig:nm3}.
The cosmological parameters were set to
$\Omega_{\Lambda}=0.7$,  $\Omega_{\text{CDM}}=0.25$,
$\Omega_{\text{Baryons}}=0.05$
and $h=65$ km sec${}^{-1}$ Mpc${}^{-1}$. With this choice of parameters,
the conformal time today is $\eta_0 \approx 1.5 \times 10^{4}$~Mpc
(we use units in which the speed of light and the scale factor at
present time are set to $1$).

{}For $n \ge -1$, the angular power spectra of TB and EB
correlations are dominated by the cutoff. In Fig.~\ref{fig:n2}, for which
we have taken $n=2$, the position of the main peak in the plots of $(1/2\pi)
l(l+1)C_l^{TB}$ and $(1/2\pi) l(l+1)C_l^{EB}$ vs $l$ corresponds to
the cutoff value of $2\pi/k_*=288$ Mpc or $k_* \eta_0 \approx 330$.
The dotted lines illustrate that
spectra diverge at large $l$ in the absence of the cutoff.

As the value of $n$ is decreased, the correlations gradually become
cutoff-independent. In Fig.~\ref{fig:nm2} we plot the spectra for
$n=-2$ with a cutoff at $2\pi/k_* = 288$ Mpc and without a cutoff.
As can be seen from the plot, above a certain scale ($l \lesssim
300$) the spectra do not depend on the cutoff and exhibit
a potentially interesting peak structure. However, spectra still
diverge on smaller scales in the absence of a cutoff.

Clearly, any constraint imposed by the future data on $v_0$ and $k_*$
will be $n$-dependent. In particular, with the chosen form of the
initial power spectrum [Eq.~(\ref{odd_{s}pectrum})],
for $n \ge -2$ a measurement
of $C_l^{TB}$ or $C_l^{EB}$ would not constrain the helical flow at
all. For example, if we optimistically assume that
$(1/2\pi) l(l+1)C_l^{TB} \sim 10^{-10}$, then for $n=2$ and
$2\pi /k_* = 288$ Mpc we obtain that $v_0 \lesssim 10^{17}$,
which is not a very useful bound. For $n=-3$ and $n=-4$, with the
same value of $2\pi /k_*$, we find
$v_0 \lesssim 10^{-5}$ and $v_0 \lesssim 10^{-11}$ respectively.

Causality will, in general, constrain the value of the spectral index
$n$ from below. The bound is obtained by setting the real space velocity
correlation function to zero at causally prohibited separations. This
constrains the Fourier transform of the correlator to be an analytical
function of $k$, which, in turn, implies that $n \ge 2$ for small $k$.
Thus, our analysis suggests that the CMBR will not be able to
constrain primordial kinetic helicity unless acausal physics is
responsible for producing them. Since models that best fit the CMBR
temperature power spectrum do rely on causality being violated in the past,
by \emph{e.g.}~inflation or a larger speed of light, it is not
inconceivable that the velocity correlations would be acausal as well.

\section{Kinetic helicity from magnetic helicity?}
\label{hvelbhel}

In this section we explore the possibility that helical
magnetic fields may induce kinetic helicity ${\bm \omega}$ at last
scattering. As we shall see, the induced velocities
are insignificant and can be ignored.

We shall be interested in the effects of a statistically
homogeneous and isotropic magnetic field, with possibly non-vanishing
helicity. If we denote the Fourier amplitudes of the magnetic field by
${\bf b}({\bf k})$, then, as in Sec.~\ref{hvel},
\begin{eqnarray}
&&
\langle b_i ({\bf k} ) b_j ({\bf k}' ) \rangle =
(2\pi )^3 \delta^{(3)}({\bf k} + {\bf k}' )
\nonumber \\
&&
\times [ (\delta_{ij} - {\hat k}_i {\hat k}_j) S(k) +
i \varepsilon_{ijl} {\hat k}_l A(k) ] .
\label{bcorr}
\end{eqnarray}
Here $S(k)$ denotes the symmetric part and $A(k)$ the
antisymmetric part of the correlator. These functions
are constrained by \cite{Frietal75}
\begin{equation}
S({\bf k}) \ge  |A({\bf k})|
\end{equation}
and $S({\bf k}) \ge 0$,  exactly as in the case of kinetic helicity
[Eq.~(\ref{maxkinhel})]. Furthermore,
$$
\langle {\bf B} ({\bf x}) \cdot
[\nabla \times {\bf B}({\bf x})] \rangle
$$
only depends on $A(k)$ and not on $S(k)$. Therefore
$A(k)$ represents the helical component of the magnetic
field and $S(k)$ the non-helical component.

The Lorentz force ${\bf F}_L$ due to the magnetic field
on the (electrically neutral) cosmic plasma is
\begin{equation}
{\bf F}_L = {\bf j}\times {\bf B} ,
\end{equation}
where ${\bf j}$ is the electric current that satisfies Maxwell's equation
in the MHD approximation:
\begin{equation}
{\bf j} = \nabla \times {\bf B} \, .
\end{equation}
Therefore
\begin{equation}
\langle {\bf F}_L \rangle =
\langle {\bf B} ({\bf x}) \times
[ \nabla \times {\bf B}({\bf x}) ] \rangle .
\end{equation}
In the tight-coupling approximation, in which it is assumed that velocities
of photons, electrons and protons are approximately the same, the Lorentz
force induces flows of neutral plasma.\footnote{In the tight-coupling
approximation the photons, electrons and protons are treated as a
single fluid. However, the current $\bf j$ is
proportional to the difference in velocities of electron and proton fluids.
It is the Lorentz force due to this slight difference in velocities that
drives flows in neutral plasma.}
An evaluation shows that the quantity $\langle {\bf F}_L \rangle$
depends on $S(k)$ but has no dependence on $A(k)$.
In other words, the term with $A(k)$ denotes the ``force-free'' component
of the magnetic field. Therefore the velocity flow at last scattering is
unaffected by the helical component and the corresponding Doppler signature
on the CMBR can only carry information about $S(k)$ and no information
about $A(k)$.

In the preceding discussion, based on the tight-coupling approximation, we
assumed that the Lorentz force acts on an element of the neutral plasma and
changes its velocity. In reality the coupling of photons to electrons is
much stronger than that to protons and so the plasma at recombination is
better treated as composed of two fluids, namely the electron-photon fluid
and the proton fluid. It is precisely in this approximation that the
generation of magnetic fields due to cosmic vorticity was analyzed by
Harrison \cite{Har70}.

Our situation is similar to Harrison's, except for initial conditions: we
have an initial (force-free) helical magnetic field and we need to find the
velocity induced by it. The analysis, described in Appendix \ref{appendixB},
closely follows that of Harrison. The result is that the electron-photon
fluid will gain a vorticity ${\bm \omega}_e$ given by
\begin{equation}
{\bm \omega}_e = {1\over {en_e}} \nabla^2 {\bf B} ,
\end{equation}
where $e$ is the electron charge and $n_e$ is the
electron number density. If we estimate
$|\nabla^2 {\bf B}| \sim B/L^2$, where $L$ is the
coherence scale of the field, we find
\begin{equation}
|{\bf v}| \sim |L {\bm \omega}_e|
\sim 10^{-18} \left ( {{B_0}\over {10^{-9}{\rm G}}} \right )
\left ( {{1 {\rm kpc}} \over {L_0}} \right ) ,
\end{equation}
where $B_0$ and $L_0$ denote the magnetic field strength
and coherence scale at the present epoch and the cosmic electron
number density is $\sim 10^{-6}/{\rm cm}^3$. (Current bounds
on cosmic magnetic fields constrain the field strength to
be less than $\sim 10^{-9}$ G.) Compared to the
velocities induced by gravitational perturbations ($\sim 10^{-5}$)
the velocities induced by helical fields are insignificant.

\section{A strategy to detect magnetic helicity}
\label{strategy}

In the previous section we have shown that only the non-helical
component of the magnetic field can have a signature
in the Doppler contribution to the CMBR. If we could
find another observable that is sensitive to both the
non-helical and the helical components, we could combine
observations and extract the helical component of the
magnetic field. An observable that does depend on both
helical and non-helical components is the Faraday rotation
of linearly polarized sources due to light propagation
through a magnetized plasma.

The CMBR is expected to be linearly polarized and so any
intervening magnetic fields will rotate the polarization vector
at a rate given by:
\begin{equation}
d\theta = 
\lambda^2 {{e^3}\over {8\pi^2 m_e^2}} a \ n_e {\bf B}\cdot d{\bf l} \ ,
\label{theta1}
\end{equation}
where $\lambda$ is the wavelength of light, 
$a$ is the scale factor normalized so that 
$a_{\text{today}}=1$, $n_e$ is the number density of free electrons, 
$d {\bf l}$ is the comoving
length element along the photon trajectory from the 
source to the observer and we
are using natural units with $\hbar=c=1$ and $\alpha=e^2/(4\pi)\approx 1/137$.
Using the known expression for Thomson scattering cross-section,
\begin{equation}
\sigma_T = {8 \pi \alpha^2 \over 3 m_e^2} \ ,
\end{equation}
and integrating along the line of sight, we obtain from Eq.~(\ref{theta1}):
\begin{equation}
\theta = {3 \over {2 \pi e}} \lambda_0^2 
\int \dot{\tau}({\bf x}) \ {\bf {\tilde B}} \cdot d {\bf l}
\label{theta2}
\end{equation}
where $\dot{\tau}({\bf x}) \equiv n_e \sigma_T a$ is the differential
optical depth along the line of sight, $\lambda_0$ is the
observed wavelength of the radiation and
${\bf {\tilde B}} \equiv {\bf B} a^2$ is the ``comoving'' magnetic field.

Faraday rotation depends on the free electron density, which becomes negligible
towards the end of recombination. Therefore, the bulk of the rotation is produced during a
relatively brief period of time when the electron density is sufficiently low for
polarization to be produced and yet sufficiently high for the Faraday rotation to occur.
The average Faraday rotation (in radians) between Thomson scatterings
due to a tangled magnetic field was calculated in Ref.~\cite{Harari} and is given by
\begin{equation}
F = {3 \over 2 \pi e} { B_0 \over \nu_0^2}
\approx 0.08 \left( {B_0 \over 10^{-9}{\rm G}} \right)
\left( {30{\rm GHz} \over \nu_0} \right)^2 \ ,
\end{equation}
where $B_0$ is the current amplitude of the field and $\nu_0$ is
the radiation frequency observed today.

The amplitude of the CMB polarization fluctuations is expected to be of order $10^{-6}$,
an order of magnitude lower than that of the temperature
fluctuations. As discussed in Ref.~\cite{KosLo96},
detecting a Faraday rotation of order $1^{\rm o}$ will require a measurement
which is superior in sensitivity by another factor of $10^{2}$. Such accuracy is at the
limit of current experimental proposals but there is a hope that it will eventually be
accomplished.

It is usual to define the wavelength independent
``rotation measure'' (RM) as:
\begin{equation}
{\rm RM}= {3 \over {2 \pi e }} \int \dot{\tau}({\bf x})
\ {\bf {\tilde B}} \cdot d{\bf l}
\label{rmtheta2}
\end{equation}
A polarization map of the CMBR at several wavelengths will
(in principle) give $\theta$ as a function of $\lambda$ along
different directions in the sky. From this information
the rotation measure in any given direction in the sky
can be determined. Hence the polarization map of the CMBR
will also lead to a ``rotation measure map''. Then we can find
correlations of the RM:
\begin{equation}
RR' \equiv \left\langle {\rm RM}(\hat{{\mathbf n}})
{\rm RM}(\hat{{\mathbf n}}')
\right\rangle ,
\end{equation}
where ${\rm RM}(\hat{{\mathbf n}})$
is the rotation measure of the CMBR, observed
along the direction $\hat{{\mathbf n}}$.
Using Eq.~(\ref{bcorr}), this yields
\begin{equation}
RR' = \left({3 \over {2 \pi e}} \right)^2
\int {{d^3k}\over {(2\pi )^3}}
\left [ \alpha S(k) + \beta A(k) \right ] ,
\label{RR'}
\end{equation}
where
\begin{equation}
\alpha \equiv {\bf K}_1 \cdot {\bf K}_2^* -
(\hat{{\mathbf k}}\cdot {\bf K}_1)(\hat{{\mathbf k}}\cdot {\bf K}_2^*) ,
\end{equation}
\begin{equation}
\beta \equiv i \hat{{\mathbf k}}\cdot ({\bf K}_1 \times {\bf K}_2^*) \, ,
\end{equation}
\begin{equation}
{\bf K}_1 \equiv \hat{{\mathbf n}} \int d\eta \ \dot{\tau}(\eta)
e^{-i{\bf k} \cdot \hat{{\mathbf n}}\eta} ,
\label{k1}
\end{equation}
\begin{equation}
{\bf K}_2^* \equiv \hat{{\mathbf n}}' \int d\eta \ \dot{\tau}(\eta)
e^{+i{\bf k} \cdot \hat{{\mathbf n}}'\eta} .
\label{k2}
\end{equation}
Eqs.~(\ref{k1}) and (\ref{k2}) are obtained under the assumption
that effects of inhomogeneities in free electron density along different
directions on the sky are of the next order in perturbation theory.
That allows us to write $\dot{\tau}({\bf x})=\dot{\tau}(\eta)$ and
$d{\bf l}=\hat{{\mathbf n}}d\eta$.

A crucial feature of $RR'$ is that it depends on both
the helical and non-helical spectral functions $S(k)$ and
$A(k)$.

The CMBR polarization spectra induced by tangled magnetic fields have
already been calculated by Seshadri and Subramanian \cite{SesSub01}.
They computed the correlator between the B-type polarization of the
CMBR photons commonly denoted by $C_l^{BB}$. Defining
$$\Delta T^{BB} (l)= T_0 \sqrt{\frac{l(l+1)}{2\pi} C_l^{BB}} ,$$
where
$T_0$ is the CMBR temperature, they found (Eq.~(8) in
Ref.~\cite{SesSub01}) that for $l>1500$,
\begin{equation}
\Delta T^{BB} (l) \simeq (0.93 \mu {\rm K} )
I\left ( {l\over {R_*}} \right )
\left [ {{B_{\text{rms}}}\over {3.10^{-9} {\rm G}}} \right ]^2
\left [ {{l}\over {1500}} \right ]^{-1/2} .
\label{DeltaTBB}
\end{equation}
Here $B_{\text{rms}}$ is the rms magnetic field strength at the present
epoch and $R_* \equiv \eta_0 - \eta_*$ is the conformal time interval
from last scattering to the present epoch. The spectral functions
$S(k)$ are contained in the mode-coupling integral $I(l/R_*)$ as
follows:
\begin{eqnarray}
I^2 (k) &=& \int_0^\infty {{dq}\over q} \int_{-1}^{+1} d\mu
{{h(q) h(|{\bf k}+{\bf q}|) k^3}\over
{|{\bf k}+{\bf q}|^3}}
\nonumber \\
&\times& (1-\mu^2 ) \left [ 1 +
{{(k+2q\mu )(k+q\mu )} \over {|{\bf k}+{\bf q}|^2}}
\right ] ,
\label{isquared}
\end{eqnarray}
where $|{\bf k}+{\bf q}| = (k^2+q^2+2kq\mu )^{1/2}$ and
\begin{equation}
h(k) \equiv {{k^3 S(k)} \over {\pi^2 B_{\text{rms}}^2}}  \ .
\end{equation}

The exact form of the Seshadri and Subramanian's result is
not important for describing our strategy to isolate the
helical component of the magnetic field. We need only observe
the crucial feature that $\Delta T^{BB}(l)$, and hence
$C_l^{BB}$, depends on the non-helical spectral function
$S(k)$ and is independent of the helical spectral function
$A(k)$.
This is because, as discussed in the previous section,
$A(k)$ is the force-free component of the magnetic field
and does not induce any velocity in the last scattering
surface. However, it is worth noting a few assumptions
that enter the analysis in Ref.~\cite{SesSub01}. The
first is an assumption of Gaussianity by which 4-point
functions of the magnetic field can be factored into
a product of 2 point functions. The second is the
assumption that the effects of Faraday rotation at
last scattering are not important. The latter assumption
is justified at the high frequencies of observation
planned for the Planck satellite \cite{SesSub01}.

Hence, if we could use $C_l^{BB}$ to obtain $S(k)$ --
which would only be possible assuming some functional
form (such as a power law) for $S(k)$ since $S(k)$ occurs
within some integrals in Eq.~(\ref{isquared}) -- we could insert
the result in the expression for $RR'$ given in Eq.~(\ref{RR'}).
This will isolate $A(k)$ in Eq.~(\ref{RR'})
and, with some assumptions about the functional form of $A(k)$,
the cosmic magnetic helicity can be evaluated.
Even if
the assumptions in the derivation of Eq.~(\ref{DeltaTBB}) are
not completely valid, the RM correlator in Eq.~(\ref{RR'})
and $C_l^{BB}$ have different dependencies on $S(k)$ and
$A(k)$, and hence the two observations can be used to
disentangle these two spectral functions.

\section{Conclusions}
\label{conclusions}

We have analyzed certain P- and CP-violating signatures in the CMBR.
If there is kinetic helicity at last scattering, it
would imprint a signature in the cross-correlators $C_l^{TB}$
and $C_l^{EB}$. Kinetic helicity can be induced by
helical magnetic fields but the effect is too small to be
significant since the helical component of magnetic fields
is force-free. Instead we have proposed another strategy for
detecting the helicity of primordial magnetic fields using
polarization and rotation measure maps of the CMBR.

The helical magnetic fields produced during electroweak
baryogenesis ($\sim 10^{-13}$~G at last scattering) are several
orders weaker than current upper bounds on the magnetic field
strength ($\sim 10^{-6}$~G) at last scattering \cite{Vac01a}.
Therefore the detection of electroweak fields does not seem
feasible with forthcoming experiments.
However, it is conceivable
that stronger helical fields were produced due to some other
mechanism and so it is still important to think of strategies
for detecting primordial magnetic fields and helicity.

\begin{acknowledgments}	
TV is grateful to George Field and Kandu Subramanian
for discussions. LP acknowledges helpful conversations with Luis Mendes.
This work was supported by DoE grant number DEFG0295ER40898 at CWRU.
LP is supported by PPARC. SW is supported by the NSF.
LP and TV would like to thank the organizers of the July 2001 ESF
COSLAB Programme workshop at Imperial College, London,
during which this project was started.
\end{acknowledgments}

\appendix

\section{}
\label{appendixA}

Here we outline our calculation of the signature of
kinetic helicity in the CMBR.

In the ``total angular momentum'' formalism of
Hu and White \cite{HuWhi97},
the observable CMB anisotropies $T, Q,$ and $U$ are
expanded into a set of components $\Theta _{l}^{(m)}$, $E_{l}^{(m)}$,
$B^{(m)}_{l}$, which are functions of $k$ and $\eta $; here
$k$ is the wavenumber in Fourier space and $\eta $ is the
conformal time. These quantities are such that the power spectra
$C_{l}^{XY}$ are expressed through the $l$-th components only
(see Eq.~(56) of \cite{HuWhi97}), \emph{e.g.}
\begin{equation}
C_{l}^{TB}=\frac{2}{\pi }\int dkk^{2}\sum _{m=-2}^{2}\frac{\left[ \Theta
^{(m)}_{l}\left( k,\eta _{0}\right) \right]
^{*}}{2l+1}\frac{B^{(m)}_{l}\left( k,\eta _{0}\right) }{2l+1},
\end{equation}
\begin{equation}
C_{l}^{EB}=\frac{2}{\pi }\int dkk^{2}\sum _{m=-2}^{2}\frac{\left[
E^{(m)}_{l}\left( k,\eta _{0}\right) \right]
^{*}}{2l+1}\frac{B^{(m)}_{l}\left( k,\eta _{0}\right) }{2l+1} \, ,
\end{equation}
where $m=0$, $m=\pm 1$ and $m=\pm 2$ denote scalar, vector and tensor
contributions respectively, and $\eta_0$ is the conformal time today.
The quantities $\Theta _{l}^{(m)}$, $E_{l}^{(m)}$, $B^{(m)}_{l}$,
in turn, are found from the linearized Einstein and Boltzmann equations
and are expressed as integrals along the line-of-sight of primordial
perturbation sources (initial density and velocity perturbations).
Assuming that the perturbation sources are random fields with known correlations in
Fourier space, one can find the expectation value of any
quadratic combination of $T$, $Q$ and $U$ or, equivalently, power spectra
$C_{l}^{TT,EE,BB}$ and cross-correlators $C_{l}^{TE,TB,EB}$.

Scalar perturbations do not generate $B$-type polarization.
We will also assume that there are no parity violating tensor sources.
The integral solution of the Boltzmann equations for the components
$\Theta _{l}^{(\pm 1)}$, $E_{l}^{(\pm 1)}$, $B_{l}^{(\pm 1)}$ is
given by Eqs.~(79) and (77) of Ref.~\cite{HuWhi97},
\begin{eqnarray}
&&\frac{\Theta_{l}^{(\pm 1)}(\eta _{0},k)}{2l+1}
\nonumber \\
&&=\int _{0}^{\eta _{0}}d\eta \, e^{-\tau }
\Big[ \dot{\tau }\big( v^{(\pm 1)}-V\big)
j_{l}^{1,\pm 1}\left( k ( \eta _{0}-\eta) \right)
\nonumber \\
&&+ \big( \dot{\tau } P^{(\pm 1)} + \frac{kV}{\sqrt{3}} \big)
j_{l}^{2,\pm 1}\left( k (\eta _{0}-\eta) \right)
\Big],
\label{lostheta}
\end{eqnarray}
\begin{eqnarray}
\frac{E_{l}^{(\pm 1)}\left( \eta _{0},k\right) }{2l+1} &=&
-\sqrt{6}\int_{0}^{\eta _{0}}d\eta \, \dot{\tau} e^{-\tau }
\nonumber \\ &\times& P^{(\pm 1)}\left(k,\eta \right)
\varepsilon_{l}^{(\pm 1)}\left(k(\eta _{0}-\eta)
\right) ,
\label{losE}
\end{eqnarray}
\begin{eqnarray}
\frac{B_{l}^{(\pm 1)}\left( \eta _{0},k\right) }{2l+1} &
= & -\sqrt{6}\int _{0}^{\eta _{0}}d\eta \, \dot{\tau} e^{-\tau }
\nonumber \\
&\times& P^{(\pm 1)}\left( k,\eta \right)
\beta_{l}^{(\pm 1)}\left( k( \eta _{0}-\eta) \right) ,
\label{losB}
\end{eqnarray}
where $V$ is the vector component of the metric perturbations,
$\dot{\tau}$ is the differential optical depth,
$\tau(\eta)=\int_{\eta}^{\eta_0}\dot{\tau}(\eta') d\eta'$,
and
\begin{equation}
P^{(\pm 1)}\left(k,\eta \right) \equiv
{1 \over 10}\left( \Theta_2^{(\pm 1)} - \sqrt{6}E_2^{(\pm 1)} \right) \, .
\label{p_def}
\end{equation}
The functions $j_{l}^{1,\pm 1}$, $j_{l}^{2,\pm 1}$,
$\varepsilon ^{(\pm 1)}_{l}$
and $\beta ^{(\pm 1)}_{l}$ satisfy $j_{l}^{11}=j_{l}^{1,-1}$,
$ j_{l}^{21}=j_{l}^{2,-1}$, $\varepsilon_{l}^{(1)}=\varepsilon _{l}^{(-1)}$
and $\beta _{l}^{(1)}=-\beta_{l}^{(-1)}$, where
\begin{equation}
j_{l}^{11}\left( x\right) \equiv \sqrt{\frac{l\left( l+1\right)
}{2}}\frac{j_{l}\left( x\right) }{x},
\end{equation}
\begin{equation}
j_{l}^{21}\left( x\right) \equiv
\sqrt{\frac{3l\left( l+1\right) }{2}}\frac{d}{dx}
\left( \frac{j_{l}\left( x\right) }{x}\right) ,
\end{equation}
\begin{equation}
\varepsilon _{l}^{(1)}\left( x\right) \equiv
\frac{\sqrt{\left( l-1\right) \left( l+2\right) }}{2}
\left( \frac{j_{l}'\left( x\right) }{x}+
\frac{j_{l}\left( x\right)} {x^{2}}\right) ,
\end{equation}
\begin{equation}
\beta _{l}^{(1)}\left( x\right) \equiv
\frac{\sqrt{\left( l-1\right) \left(
l+2\right) }}{2}\frac{j_{l}\left( x\right) }{x},
\end{equation}
and $j_{l}\left( x\right) $ is the spherical Bessel function.
It is easy to see that, for instance, the correlator $C_{l}^{EB}$
depends on the parity-odd combination $P^{(1)}P^{(1)}-P^{(-1)}P^{(-1)}$.

The CMBR anisotropies sourced by velocity flows are predominantly
due to the Doppler effect. Since the net effect is expected to be
small, it is a good approximation to keep only terms that are of
lowest non-trivial order in $v^{(\pm 1)}$ and in the
tight coupling approximation. Then Eq.~(\ref{lostheta}) becomes
\begin{eqnarray}
\frac{\Theta^{(\pm 1)}_{l}\left( \eta _{0},k\right) }{2l+1}&=
&
\int _{0}^{\eta_{0}} d\eta \,
\dot{\tau} e^{-\tau } 
\nonumber \\
&
\times
&
v^{(\pm 1)}( k,\eta )
j_{l}^{1,\pm 1}( k( \eta _{0}-\eta
) ) .
\label{doppler}
\end{eqnarray}
One can express functions
$P^{(\pm 1)}$ in terms of $v^{(\pm 1)}$ using the
linearized Boltzmann equations for temperature and polarization
anisotropies written in form of infinite recursive series in multipole
index $l$ (see equations (60), (63) and (64) of \cite{HuWhi97}).
The relevant equations are:
\begin{equation}
\dot{\Theta}^{(\pm 1)}_{1}=-k\frac{\sqrt{3}}{5} \Theta^{(\pm 1)}_{2} -
\dot{\tau} ( \Theta^{(\pm 1)}_{1} - v^{(\pm 1)} - \dot{V} )
\label{boltz_1} \, ,
\end{equation}
\begin{equation}
\dot{\Theta}^{(\pm 1)}_{2}= k\left[\frac{1}{\sqrt{3}} \Theta^{(\pm 1)}_{1} -
\frac{2\sqrt{2}}{7} \Theta^{(\pm 1)}_{3} \right] -
\dot{\tau} ( \Theta^{(\pm 1)}_{2} - P^{(\pm 1)} ) \, ,
\label{boltz_2}
\end{equation}
\begin{equation}
\dot{E}^{(\pm 1)}_{2}= - k \frac{\sqrt{40}}{7\sqrt{9}} E^{(\pm 1)}_{3}
- \dot{\tau} ( E^{(\pm 1)}_{2} + \sqrt{6} P^{(\pm 1)} ) \, .
\label{boltz_3}
\end{equation}
If vector metric perturbations are exclusively due to
velocity flows, then $\dot{V}$ is of second order in $v^{\pm 1}$ and
its effect on CMBR photons is small compared to the induced Doppler shifts.
Tight coupling implies that the mean free path of photons is
negligible compared to the scales under consideration:
$\dot{\tau}^{-1} \to 0$ and $k/\dot{\tau} \ll 1$.
{}From Eq.~(\ref{boltz_1}), to $0$-th order in the
tight coupling approximation, we obtain
$\Theta_1^{(\pm 1)}=v^{(\pm 1)}$. To the
same order, from Eqs.~(\ref{boltz_3}) and (\ref{p_def}), we find that
\begin{equation}
E_2^{(\pm 1)}= - {\sqrt{6} \over 4} \Theta_2^{(\pm 1)} \ .
\label{ETheta}
\end{equation}
The quadrupole moments
$\Theta_2^{(\pm 1)}$ and $P^{(\pm 1)}$ vanish when $\dot{\tau}^{-1} \to 0$.
However, a non-zero dipole term generates the quadrupole moments
in the next lowest order in $k/\dot{\tau}$.
Using Eqs.~(\ref{p_def}) and (\ref{ETheta}),
Eq.~(\ref{boltz_2}) can be re-written as:
\begin{equation}
\Theta_2^{(\pm 1)}-{4 \over 3} \dot{\tau}^{-1 }\dot{\Theta}^{(\pm 1)}_{2}+
{4k\over 3\dot{\tau}}\frac{2\sqrt{2}}{7} \Theta^{(\pm 1)}_{3}
= {4k\over 3\sqrt{3}\dot{\tau}} \Theta^{(\pm 1)}_{1}.
\end{equation}
In the above, the RHS is the source, namely, each of the terms on the
LHS would be zero in the absence of $\Theta_1^{(1)}$. On the LHS, the
leading order (in tight coupling) term is $\Theta_2^{(\pm 1)}$. Therefore,
$\Theta_2^{(\pm 1)} \approx 4 k v^{(\pm 1)}/(3\sqrt{3}\dot{\tau})$ and
\begin{equation}
P^{(\pm 1)}\approx \frac{kv^{(\pm 1)}}{3\sqrt{3}\dot{\tau}} \, ,
\label{polsource}
\end{equation}

{}From Eqs.~(\ref{losE}), (\ref{losB}), (\ref{doppler}) and (\ref{polsource})
we obtain the following expressions for the
spectral correlators $C_{l}^{TB}$ and $C_{l}^{EB}$:
\begin{eqnarray}
\label{uetc_tb}
&C_{l}^{TB}=-\frac{2\sqrt{2}}{3\pi }\int _{0}^{\infty }k^{2}dk\, \, k
\int _{0}^{\eta _{0}}d\eta _{1}\int _{0}^{\eta _{0}}d\eta_{2}&
\\
&\times
e^{-\tau\left(\eta _{1}\right)}e^{-\tau\left(\eta _{2}\right)}
\varepsilon _{l}^{(1)}\left( k\left( \eta _{0}-\eta_{1}\right) \right)
j_{l}^{11}\left( k\left( \eta _{0}-\eta _{2}\right) \right)&
\nonumber \\
&\times \langle
v^{(1)}\left( k,\eta _{1}\right) v^{(1)}\left( k,\eta_{2}\right)
-v^{(-1)}\left( k,\eta _{1}\right) v^{(-1)}\left(
k,\eta _{2}\right)
\rangle \, . &
\nonumber
\end{eqnarray}
\begin{eqnarray}
\label{uetc_eb}
&C_{l}^{EB}=\frac{4}{9\pi }\int _{0}^{\infty }k^{2}dkk^{2}
\int _{0}^{\eta _{0}}d\eta _{1}\int _{0}^{\eta _{0}}d\eta_{2}&
\\
&\times
e^{-\tau\left(\eta _{1}\right)} e^{-\tau\left(\eta _{2}\right)}
\varepsilon_{l}^{(1)}\left(k(\eta_{0}-\eta_{1}) \right)
\beta _{l}^{(1)}\left( k(\eta_{0}-\eta_{2}) \right)&
\nonumber \\
&\times \langle v^{(1)}\left( k,\eta_{1}\right)
v^{(1)}\left(k,\eta_{2}\right)-v^{(-1)}\left(k,\eta_{1}\right)
v^{(-1)}\left(k,\eta_{2}\right) \rangle . &
\nonumber
\end{eqnarray}
If we assume that the time-evolution of each Fourier
mode of the velocity field is independent of $\hat{k}$, \emph{i.e.}~the
evolution equations contain only $k=|{\bf k}|$, then we can write
the unequal time correlator in Eqs.~(\ref{uetc_tb}) and (\ref{uetc_eb})
as a product of the initial power spectrum and the evolution functions
${\cal T}(k,\eta)$:
\begin{eqnarray}
\label{split}
&&\langle v^{(1)}( k,\eta_{1}) v^{(1)}(k,\eta_{2})-
v^{(-1)}(k,\eta_{1}) v^{(-1)}(k,\eta_{2}) \rangle
\nonumber \\
&=&{\cal T}(k,\eta_1) {\cal T}(k,\eta_2)
\nonumber \\
&\times & \langle v^{(1)}(k) v^{(1)}(k) -
v^{(-1)}(k) v^{(-1)}(k) \rangle ,
\end{eqnarray}
where
$\langle v^{(1)}(k) v^{(1)}(k) -
v^{(-1)}(k) v^{(-1)}(k) \rangle$
is the spectrum evaluated at some initial
time which we choose to be the time of recombination.
Using Eq.~(\ref{odd_{s}pectrum}) for the velocity power spectrum together
with Eqs.~(\ref{uetc_tb}-\ref{split}), we can
write $C_{l}^{TB}$ and $C_{l}^{EB}$ as
\begin{equation}
C_{l}^{TB}=\frac{2}{\pi }\int
_{0}^{\infty }k^2 dk v_{0}^{2} \frac{k^n}{k_{*}^{n+3}} g(k)
\tilde{\Delta}_{l}(k)\tilde{B}_{l}(k) ,
\label{TBexact}
\end{equation}
\begin{equation}
C_{l}^{EB}=\frac{2}{\pi }\int
_{0}^{\infty }k^2 dk v_{0}^{2}
\frac{k^n}{k_{*}^{n+3}} g(k) \tilde{E}_{l}(k)\tilde{B}_{l}(k) ,
\label{EBexact}
\end{equation}
where
\begin{eqnarray}
\tilde{\Delta }_{l}(k) & \equiv  & \int _{0}^{\eta _{0}}d\eta e^{-\tau
}\dot{\tau }\, \, j^{11}_{l}(k(\eta _{0}-\eta )){\cal T}(k,\eta )
\nonumber \\ & = & \sqrt{\frac{l(l+1)}{2}}
\int _{0}^{\eta _{0}}d\eta \, \,
\frac{j_{l}(k(\eta _{0}-\eta ))}{k(\eta _{0}-\eta )} \nonumber \\
&&
\quad \quad \quad \quad \quad \quad
\times {\cal T}(k,\eta )\dot{\tau}e^{-\tau } \, \, ,
\end{eqnarray}
\begin{eqnarray}
\tilde{E}_{l}(k) & \equiv  &
- \frac{\sqrt{2}}{3}\int _{0}^{\eta _{0}}d\eta e^{-\tau }\, \,
\varepsilon ^{(1)}_{l}(k(\eta _{0}-\eta )){\cal T}(k,\eta )\nonumber \\ & = &
-\frac{\sqrt{(l-1)(l+1)}}{3\sqrt{2}}\int _{0}^{\eta _{0}}d\eta
\frac{j_{l}(k(\eta_{0}-\eta ))}{k(\eta _{0}-\eta )}
\nonumber \\ && \quad \quad \quad \quad \quad \quad \quad
\times \frac{\partial }{\partial \eta }\left(
e^{-\tau }{\cal T}(k,\eta )\right) \, ,
\end{eqnarray}
\begin{eqnarray}
\tilde{B}_{l}(k) & \equiv  &
-\frac{\sqrt{2}}{3}\int _{0}^{\eta _{0}}d\eta e^{-\tau }\, \,
\beta ^{(1)}_{l}(k(\eta _{0}-\eta )){\cal T}(k,\eta )\nonumber \\ & = &
-\frac{\sqrt{(l-1)(l+1)}}{3\sqrt{2}}
\int _{0}^{\eta _{0}}d\eta \frac{j_{l}(k(\eta
_{0}-\eta ))}{\eta _{0}-\eta }
\nonumber \\ &&
\quad \quad \quad \quad \quad \quad \quad
\times {\cal T}(k,\eta )e^{-\tau }.
\end{eqnarray}

As a first step, let us assume that the velocity field is
simply being red-shifted by the expansion of the universe. Then
\begin{equation}
{\cal T}(k,\eta )=\left(\frac{a_{rec}}{a} \right)^2  \, .
\label{aketa}
\end{equation}

{}For small $l$ we can evaluate $C_l^{TB}$ and $C_l^{EB}$ approximately,
assuming $g(k)=1$:
\begin{equation}
\tilde{\Delta }_{l}(k) \approx
\sqrt{\frac{l(l+1)}{2}}
\frac{j_{l}(k\eta _{0})}{k\eta _{0}} ,
\end{equation}
\begin{equation}
\tilde{E}_{l}(k) \approx
\frac{\sqrt{(l-1)(l+1)}}{3\sqrt{2}} \frac{j_{l}(k\eta_{0})}{k\eta _{0}}
\left[1+2\left({\dot{a} \over a\dot{\tau}}\right)_{rec} \right],
\end{equation}
\begin{equation}
\tilde{B}_{l}(k) \approx
- \frac{\sqrt{(l-1)(l+1)}}{3\sqrt{2}} \frac{j_{l}(k\eta_{0})}{\eta _{0}}
\dot{\tau}^{-1}_{rec} \, ,
\end{equation}
\begin{eqnarray}
C_{l}^{TB}&=&
\frac{(l+1)\sqrt{l(l-1)}}{3\pi} \frac{v_0^2}{\dot{\tau}_{rec}\eta_0}
(k_*\eta_0)^{-(n+3)} \nonumber \\ &\times&
\int _{0}^{\infty} dx \, x^{n+1} j^2_l(x) \ ,
\label{cltbfinal}
\end{eqnarray}
\begin{eqnarray}
C_{l}^{EB}&=&-\frac{(l-1)(l+1)}{9\pi} \frac{v_0^2}{\dot{\tau}_{rec}\eta_0}
(k_*\eta_0)^{-(n+3)} \nonumber \\ &\times&
\left[1+2\left({\dot{a} \over a\dot{\tau}}\right)_{rec} \right]
\int _{0}^{\infty} \!dx \, x^{n+1} j^2_l(x).
\label{clebfinal}
\end{eqnarray}
The last integral can be evaluated:
\begin{equation}
\int _{0}^{\infty} dx x^{n+1} j^2_l(x)= {{2^{n-1}\pi \Gamma(-n)
\Gamma(l+1+{n\over 2})}
\over \Gamma^2({{1-n}\over 2}) \Gamma(l+1-{n\over 2})} .
\end{equation}
We find for $n < 0$ that
\begin{equation}
C_l^{TB,EB} \propto l^{n+2} .
\end{equation}
For $n \ge 0$ the integral diverges and is dominated by the range
near the cutoff:
\begin{equation}
\int _{0}^{x_0} dx x^{n+1} j^2_l(x) \sim {x_0^{n+2} \over n+1}
\end{equation}
and $C_l^{TB,EB} \propto l^2$.

The expressions in Eqs.~(\ref{TBexact}) and (\ref{EBexact})
have been evaluated numerically and the results are described in
Sec. \ref{hvelcmbr}.

\section{}
\label{appendixB}

Here we find the vorticity induced by magnetic fields at
last scattering in the two-fluid approximation where
the photon and electron velocities are equal, but the
proton velocities can be different. This is closely analogous
to Harrison's calculation of the induced magnetic field due
to cosmic vorticity \cite{Har70}.

The starting point is the Euler flow equations for the velocities
${\bf v}_{b}$, ${\bf v}_{e}$, ${\bf v}_{\gamma }$
of the baryon (\emph{i.e.}~proton), electron and photon fluids.
These equations are all of the form
\begin{equation}
\label{eq:NL}
\rho _{i}\frac{d{\mathbf v}_{i}}{dt}=
{\bf F}_{i}-\sum _{j\neq i}{\bf P}_{ij} \, ,
\end{equation}
where
\begin{equation}
\label{eq:td}
\frac{d{\mathbf v}}{dt}=
\frac{\partial {\mathbf v}}{\partial t}+
( {\bf v}\cdot \nabla ) {\bf v}
\end{equation}
is the total derivative of the velocity,
\begin{equation}
{\mathbf F}_{i}=
n_{i}Z_{i}e( {\mathbf E}+{\mathbf v}_{i}\times {\mathbf B})
\end{equation}
is the force from the electromagnetic field per unit volume,
$Z_{i}e$ is the charge of particles constituting the fluid
$i$, and ${\mathbf P}_{ij}=-{\mathbf P}_{ji}$ is the rate
of momentum transfer to the fluid $i$ due to collisions
with the fluid $j$ {[}hence the negative sign in Eq.~(\ref{eq:NL}){]}
per unit volume. For baryons we take $Z_{b}=+1$. We can also
add a gradient of the gravitational potential $\nabla \phi $
to the equation for baryons. The density of particles is
$\rho _{i}=n_{i}m_{i}$.

Collisions between photons and baryons are disregarded because they
transfer much less momentum than collisions between other fluids.
Collisions of electrons with baryons determine the conductivity
$\sigma $ of the plasma:
\begin{equation}
{\mathbf P}_{eb}=\frac{en_{e}}{\sigma }{\mathbf j}.
\end{equation}

The velocities ${\mathbf v}_{i}$ of the fluids are almost equal to
each other due to tight coupling, however the velocities of electrons
and photons are closer to each other than the velocities of electrons
and baryons:
\begin{equation}
\left| {\mathbf v}_{\gamma }-{\mathbf v}_{e}\right|
\ll \left| {\mathbf v}_{e}-{\mathbf v}_{b}\right|
\ll {\mathbf v}_{b}.
\end{equation}
We take the velocities of the electrons and photons to be the
same, except when computing ${\mathbf P}_{e\gamma }$.

We now rewrite the Euler flow equations for baryons electrons, expressing
the combination $ {\mathbf E}+{\mathbf v}_{i}\times {\mathbf B} $
through other quantities:
\begin{equation}
\label{eq:fl-b}
{\mathbf E}+{\mathbf v}_{b}\times {\mathbf B}=
\frac{m_{b}}{e}\left( \frac{d{\mathbf v}_{b}}{dt}+
\nabla \phi \right) +\frac{1}{\sigma }{\mathbf j},
\end{equation}
\begin{equation}
\label{eq:fl-e}
{\mathbf E}+{\mathbf v}_{e}\times {\mathbf B}=
-\frac{m_{e}}{e}\frac{d{\mathbf v}_{e}}{dt}+
\frac{1}{\sigma }{\mathbf j}-\frac{1}{en_{e}}{\mathbf P}_{e\gamma }.
\end{equation}
The flow equation for photons is
\begin{equation}
\label{eq:fl-g}
\left( \rho _{\gamma }+p_{\gamma }\right)
\frac{d{\mathbf v}_{\gamma }}{dt}+
{\mathbf v}_{\gamma }\frac{dp_{\gamma }}{dt}=
-\left( \rho _{\gamma }+p_{\gamma }\right)
\nabla \phi -{\mathbf P}_{e\gamma },
\end{equation}
and we use the equation of state
$
p_{\gamma }= \rho _{\gamma }/3
$.
%


Maxwell's equations in the MHD approximation are
\begin{equation}
\nabla \times {\mathbf E}=
-\frac{\partial{\mathbf B}}{\partial t},\quad
\nabla \times {\mathbf j}=-\nabla ^{2}{\mathbf B}.
\end{equation}
Taking curl of Eq.~(\ref{eq:fl-b}),
%
%
using $\nabla\cdot {\bf B}=0$ and Eq.~(\ref{eq:td}) for the total derivative $d/dt$, we obtain
\begin{eqnarray}
&&
\frac{e}{m_b \sigma }\nabla ^{2}{\mathbf B} =
{e\over {m_b}} \left [ \frac{d{\mathbf B}}{dt}+
{\mathbf B}\left( \nabla \cdot {\mathbf v}_{b}\right) -
\left( {\mathbf B}\cdot \nabla \right) {\mathbf v}_{b}\right ]
\nonumber \\
&&
+
\nabla \times [ ( {\mathbf v}_{b}\cdot \nabla ) {\mathbf v}_{b}] -
\left( {\mathbf v}_{b}\cdot \nabla \right) {\bm \omega} _{b}
+ \frac{d{\bm \omega}_{b}}{dt} .
\label{eq:first-big}
\end{eqnarray}
Next we write
\begin{equation}
{\bf v}_b = {\bf u} + {\bf w},
\end{equation}
where
${\bf u} = {\dot R} {\bf r}/R$ is the velocity due to Hubble
expansion, $R(t)$ is the universal scale factor, and ${\bf w}$
is the vortical velocity in the angular
direction around the line of vorticity which is taken to coincide
with the direction of the magnetic field lines (assumed to be
parallel). Then, after some manipulations, we obtain
%
%
\begin{equation}
\frac{m_{b}}{e}\left( \frac{d{\bm \omega} _{b}}{dt}+
2{\bm \omega} _{b}\frac{\dot{R}}{R}\right) +
\frac{d{\mathbf B}}{dt}+2{\mathbf B}\frac{\dot{R}}{R}=
\frac{1}{\sigma }\nabla ^{2}{\mathbf B}.
\end{equation}
%
%
%
%
%
%
%
We can rewrite this as
\begin{equation}
\frac{d}{dt}\left( \frac{m_{b}}{e}{\bm \omega} _{b}R^2 +
{\mathbf B} R^2\right)
= \frac{R^2}{\sigma }\nabla ^{2}{\mathbf B} .
\label{eqone}
\end{equation}
%

In Eq.~(\ref{eq:fl-e}), Harrison neglects the $d{\mathbf v}_{e}/dt$
terms and also takes ${\mathbf v}_{e}={\mathbf v}_{\gamma }$ in
the ${\mathbf v}_{e}\times {\mathbf B}$ term. Taking curl of that
equation then gives
\begin{equation}
\label{eq:P1}
\frac{d{\mathbf B}}{dt}+2{\mathbf B}\frac{\dot{R}}{R}=
-\frac{1}{en_{e}}\nabla \times {\mathbf P}_{e\gamma }+
\frac{1}{\sigma }\nabla ^{2}{\mathbf B}.
\end{equation}
(Harrison also neglects the $\nabla ^{2}{\mathbf B}$ term
which we keep here.)
The unknown quantity $\nabla \times {\mathbf P}_{e\gamma }$ in
Eq.~(\ref{eq:P1}) is found by taking the curl of Eq.~(\ref{eq:fl-g}),
\begin{equation}
\label{eq:P2}
4\rho _{\gamma }\left( \frac{d{\bm \omega} _{\gamma }}{dt}+
{\bm \omega} _{\gamma }\frac{\dot{R}}{R}\right) =
-3\nabla \times {\mathbf P}_{e\gamma }.
\end{equation}
Combining Eqs.~(\ref{eq:P1}) and (\ref{eq:P2}), we get
\begin{equation}
\label{eq:P3}
\frac{d{\mathbf B}}{dt}+2{\mathbf B}\frac{\dot{R}}{R}=
\frac{4}{3}\frac{1}{en_{e}}\rho _{\gamma }
\left( \frac{d{\bm \omega} _{\gamma }}{dt}+
{\bm \omega} _{\gamma }\frac{\dot{R}}{R}\right)
+ \frac{1}{\sigma }\nabla ^{2}{\mathbf B} .
\end{equation}
Note that $\rho _{\gamma }\left( t\right) =
\rho _{\gamma }^{(0)}R^{-4}$
and $\rho _{e}\left( t\right) =\rho _{e}^{(0)}R^{-3}$.
Multiplying Eq.~(\ref{eq:P3}) by $R^{2}$ and rewriting it
as a total time derivative,
\begin{equation}
\frac{d}{dt}\left[ {\mathbf B}R^{2}-
\frac{4m_e\rho _{\gamma }^{(0)}}{3e\rho _{e}^{(0)}}
                    {\bm \omega} _{\gamma }R \right] =
\frac{R^2}{\sigma }\nabla ^{2}{\mathbf B} .
\label{eqtwo}
\end{equation}

Subtracting Eq.~(\ref{eqtwo}) from Eq.~(\ref{eqone}) and integrating
with the initial condition of zero vorticity finally gives:
\begin{equation}
{{m_b R^2}\over e} {\bm \omega}_b =
- {{4\rho_\gamma^{(0)} m_e R}\over {3e\rho_e^{(0)}}}
{\bm \omega}_\gamma \, .
\label{simeq1}
\end{equation}

Additionally, we take the curl of
\begin{equation}
\label{eq:j}
{\mathbf v}_{b}-{\mathbf v}_{e}=\frac{1}{en_{e}}{\mathbf j} \ ,
\end{equation}
equate ${\bm \omega} _{\gamma }$ and ${\bm \omega} _{e}$, and obtain
\begin{equation}
{\bm \omega} _{b}-{\bm \omega} _{\gamma }=
-\frac{1}{en_{e}}\nabla ^{2}{\mathbf B}.
\label{simeq2}
\end{equation}

Now we have two simultaneous equations, Eq.~(\ref{simeq1}) and
(\ref{simeq2}) for the two vorticities
${\bm \omega}_e={\bm \omega}_\gamma$ and ${\bm \omega}_b$.
Recognizing that $\rho_\gamma \ll \rho_b$ at last scattering,
we then solve the simultaneous equations to get
\begin{equation}
{\bm \omega}_e = {1\over {en_e}} \nabla^2 {\bf B} \, .
\label{omegaeresult}
\end{equation}

\end{document}